\documentclass[11pt,twoside]{article}
\usepackage{newpasp, psfig}
\begin{document}
\pagestyle{myheadings}

\title{XMM-Newton and Chandra view of the Cluster Abell 85 and its X-Ray Filament}

\author{Gast\~ao B. Lima Neto$^{1}$ and Florence Durret$^{2}$}

\markboth{Lima Neto \& Durret}{XMM-Newton and Chandra view of the Cluster Abell 85}

\affil{$^1$ Instituto de Astronomia,  Geof\'{\i}sica e Ci\^encias 
Atmosf\'ericas, Universidade de S\~ao Paulo, Brazil --
\rm \texttt{gastao@astro.iag.usp.br}}

\affil{$^{2}$ Institut d'Astrophysique de Paris, France -- 
\rm \texttt{durret@iap.fr}}

\begin{abstract}
    From ROSAT data, the bright X-ray cluster Abell 85 was found to show a
    roughly symmetrical shape, on to which are superimposed several features,
    among which: (1) a blob or group of galaxies falling on to the main
    cluster, with the gas in the impact region probably hotter and the
    galaxies in that zone showing enhanced star formation; (2) a filament,
    either diffuse or made of small groups of galaxies, extending at least 4
    Mpc from the cluster (Durret et al. 1998). Preliminary results obtained
    from XMM-Newton and Chandra observations of Abell 85 and its filament will
    be presented.
\end{abstract}

\section{Introduction}

The hot ($1 \la T \la 10\,$keV; central density, $n_{0}\approx 10^{-3}
${cm}${}^{-3}$) X-ray emitting gas found in rich clusters of galaxies is an
excellent tool to probe the cluster dynamics, morphology, and history.

We present here an X-ray study of Abell 85 ($z = 0.0556$, richness class
$R=0$, B-M type III) using new XMM-Newton data obtained in January 2002 and
public data available from the Chandra archive.

At the distance of Abell 85, 1 arcmin corresponds to $65 h_{70}^{-1}$kpc 
(assuming $\Omega_{M} = 0.3$ and $\Omega_{\Lambda} = 0.7$).

\section{The Data}

\begin{figure*}[!htb]
   \centering
   \mbox{\psfig{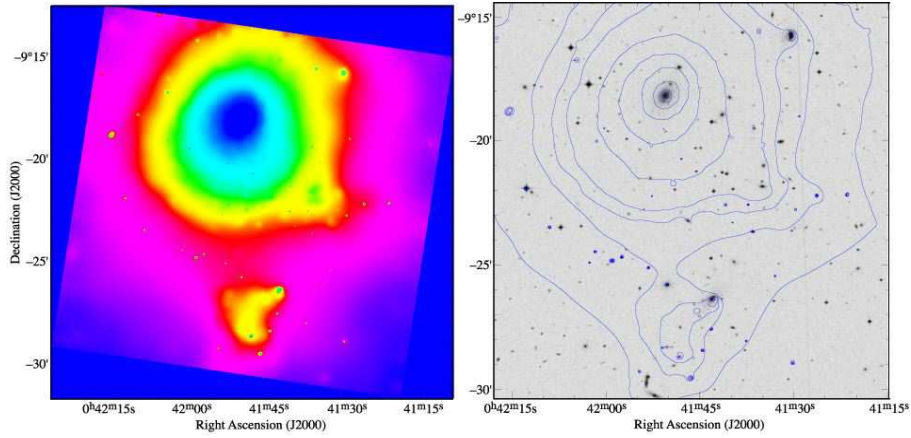}}
   \caption{Left: Flux-corrected, adaptatively smoothed X-ray image obtained
   with Chandra ACIS-I in the broad band [0.3--7.0]~keV. Right: X-ray
   isocontours superimposed on a DSS image. }
   \label{fig:a85_ACIS_I_cor}
\end{figure*}

Abell~85 was observed in august 2001 by the Chandra satellite (P.I. C.
Sarazin) with the ACIS-I detector. The net exposure time was $38.9$~ks. Figure
\ref{fig:a85_ACIS_I_cor} shows the Chandra X-ray image and the Chandra
isocontours superimposed on a DSS optical image. Notice the irregular form of
the South Blob and the excess X-ray emission at the south-west of the main
structure.

The two XMM-Newton observations were performed on January 7th,
2002 (P.I. F. Durret). The two exposures were centered on the centre
of Abell~85 (coincident with the cD galaxy) and on the south filament
respectively. Both exposure times were $12.0$~ks.

\section{The Centre, the South Blob and the Filament}

\begin{figure*}[!htb]
    \centering
    \mbox{\psfig{figure=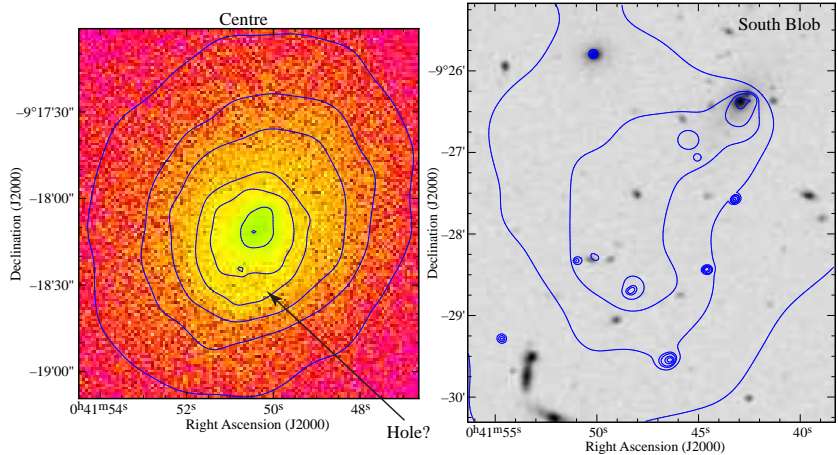,width=11cm}}
    \caption{Left: Zoom of the central region seen by Chandra ACIS-I. Right:
    The South Blob. DSS optical image superimposed with adaptatively smoothed
    Chandra ACIS-I X-ray iso-contours. Notice that the scales are different.}
    \label{fig:a85_ACIS_centreZoom}
\end{figure*}

The left panel of Figure \ref{fig:a85_ACIS_centreZoom} shows a close-up of the
central region seen by Chandra. About 15 arcsec southward of the central X-ray
maximum there seems to be a ``hole'' or ``bubble'' in the X-ray emissivity.
There is also a small X-ray point source (which is associated with an optical
counterpart) just a few arcsecs to the east of this ``hole''. Such a ``hole''
may be produced by the relativistic jets (cf. Ensslin et al 1998) either from
the central dominant galaxy or from the object associated with the X-ray point
source. Some hot bubbles have been observed in a number of clusters (Mazzotta
et al. 2002a,b; Soker et al. 2002).

The South Blob (the second brightest diffuse sub-structure) is shown in the
right panel of Figure~\ref{fig:a85_ACIS_centreZoom}. Notice that it is a
highly irregular sub-structure, with the X-ray emission dominated by a point
source (a bright galaxy at the redshift of the cluster, see also Lima Neto et
al. 1997).

\begin{figure*}[!htb]
    \centering
    \mbox{\psfig{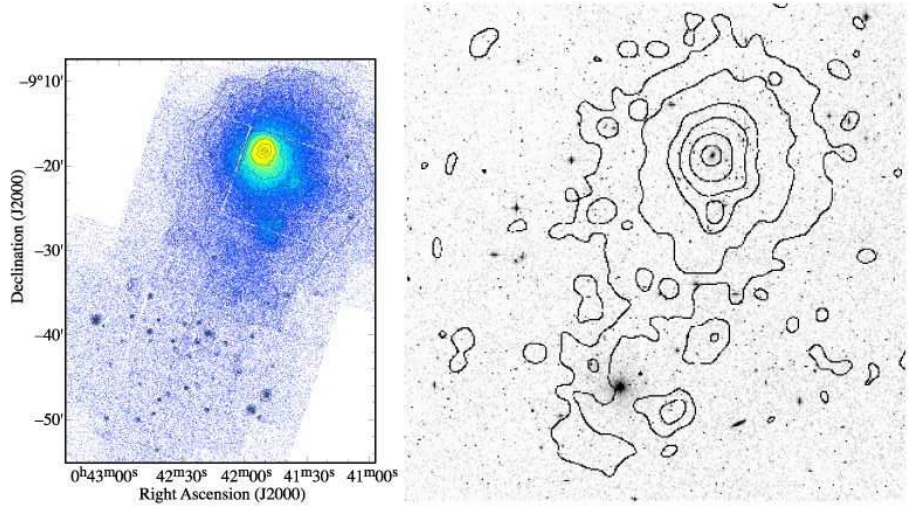}}
    \caption{Left: XMM-Newton EPIC MOS 1 and 2 mosaic
    image showing the central part of Abell 85 and its filament south of the
    cluster. Right: DSS image on to which are superimposed the smoothed ROSAT
    PSPC contours showing for the first time the X-ray filament of Abell 85
    (Durret et al. 1998). }
    \label{fig:a85_M1_tudo8000_PSPC}
\end{figure*}

\begin{figure*}[!htb]
    \centering
    \mbox{\psfig{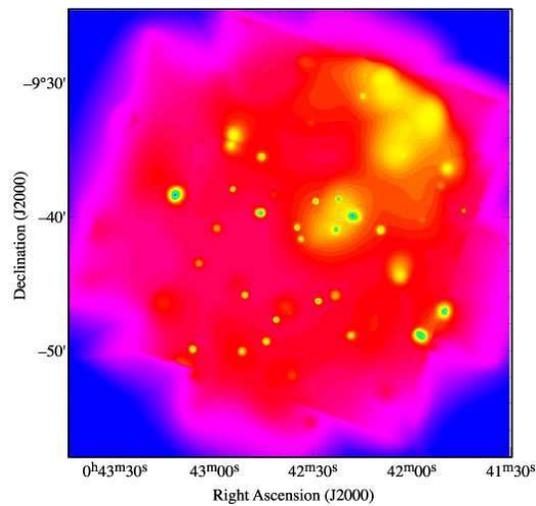}}
    \caption{Close-up of Abell 85 filament. This is an
    adaptatively smoothed XMM image combining the MOS and PN detectors.}
    \label{fig:a85_filam_M12PN_8000SMO}
\end{figure*}

Figure \ref{fig:a85_M1_tudo8000_PSPC} shows the mosaic XMM image
obtained after combining both observations. The images are the merged data
from EPIC-MOS 1 and 2 in the [0.3--8.0]~keV band. It is clear that Abell~85 has
a diffuse emission extending southward; we interpret this as part of a
filament. Figure~\ref{fig:a85_filam_M12PN_8000SMO} shows a zoom of the
filament region. Filaments are expected to connect cluster in the hierarchical 
scenario of large scale structure formation (e.g. Jenkins et al. 1998).

\section{Temperature and Metallicity Maps}

We have computed the temperature and metallicity maps from Chandra data by
dividing the central region of Abell 85 into a mesh and, for each cell,
fitting an absorbed MEKAL plasma model (Kaastra \& Mewe 1993; Liedahl et al.
1995).

\begin{figure*}[htb]
    \centering
    \mbox{\psfig{figure=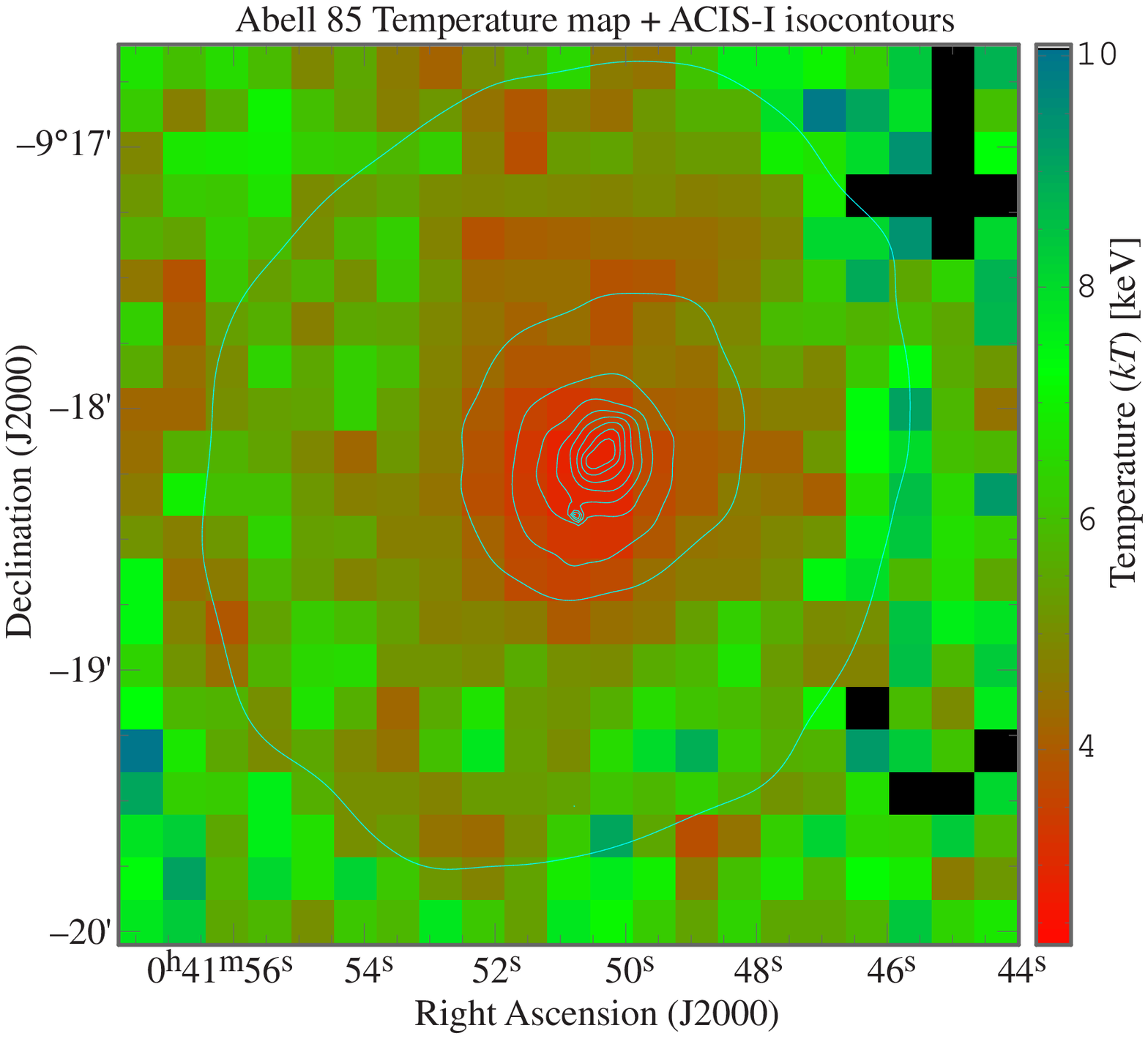,width=6.6cm}\kern6pt
    \psfig{figure=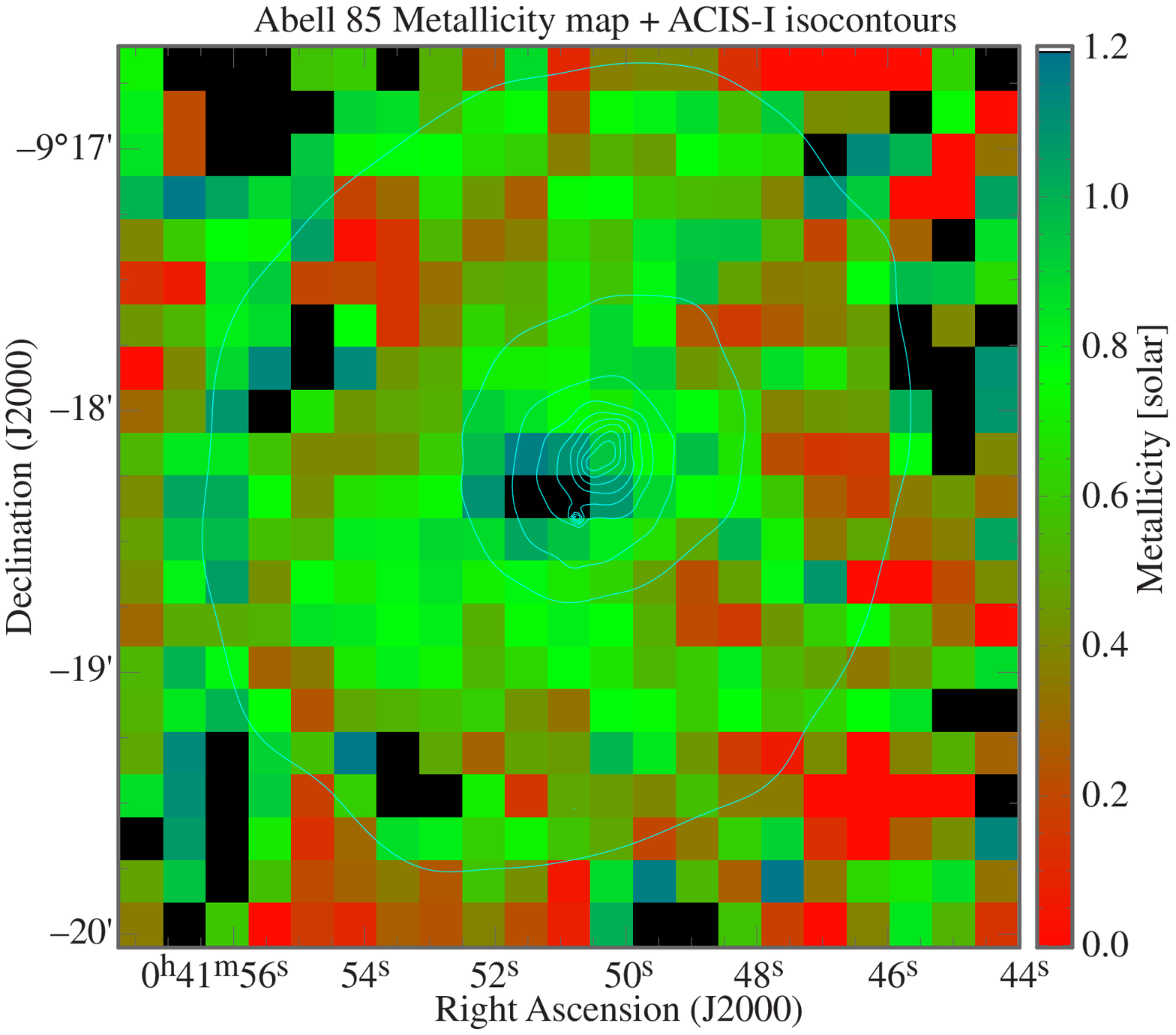,width=6.8cm}} 
    \caption{Left: Temperature map of Abell 85 central region. The X-ray
    isocontours are superimposed. Right: Metallicity map of the same region.}
    \label{fig:a85_kT_acisI2}
\end{figure*}

Figure \ref{fig:a85_kT_acisI2} shows the temperature and metallicity maps. The
X-ray isocontours are overlaid on each map. The cluster is found to be cooler
in the central regions where the metallicity is higher.

\section{Radial profiles}

\begin{figure*}[!htb]
    \centering
    \mbox{\psfig{figure=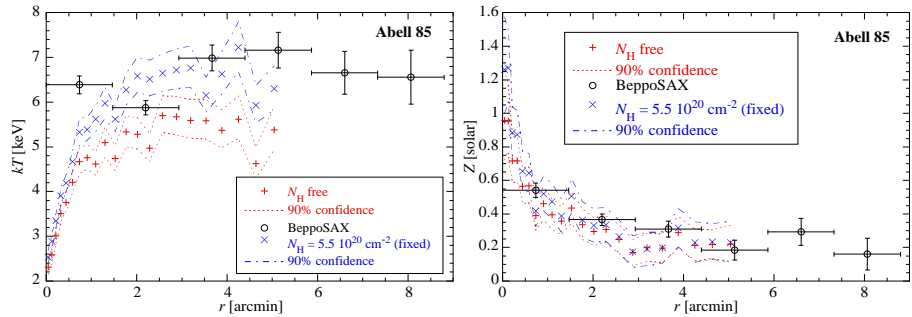,width=12cm}}
    \caption{Left: temperature profile. Right: metallicity profile. In both
    cases, the fits were done with both fixing the hydrogen column density to
    the galactic value and leaving it as a free parameter.}
    \label{fig:T_Z-R_Chandra-Beppo}
\end{figure*}

Figure \ref{fig:T_Z-R_Chandra-Beppo} shows the radial profiles of
temperature and metal abundance computed in circular rings around the cluster
centre. Notice that there is a significant difference when we fix or not the
absorbing hydrogen column density. When fixed, we used the galactic value 
given by Dickey \& Lockman (1990), available with the FTOOLS package.

We compare these profiles to those obtained by Lima Neto et al. (2001) 
using BeppoSAX data. With the low resolution of BeppoSAX, the central 
region profile is not adequately determined. With Chandra ACIS-I, there is a 
clear temperature drop towards the centre.

We also observe a steep rise in the metallicity towards the centre.

\section{Conclusions}

The high resolution and sensibility of XMM-Newton and Chandra satellite 
provides us with a new vision of an old cluster (quoting a meeting in 
Marseille five years ago...):

\begin{itemize}
    \itemsep = 0pt
 \item The X-ray emission in the central region is not totally homogeneous or
 symmetric: the maximum is displaced relatively to the outer isophotes, a
 ``hole'' seems to exist south of the center;
 
 \item The temperature is cooler and the metallicity is higher in the central
 regions of the cluster than was previous measured;

 
 \item The existence of an X-ray filament, or at least diffuse X-ray emission
 south east of the cluster is confirmed; The south blob is definitely not a
 relaxed structure.
 
\end{itemize}

\acknowledgments We acknowledge support from the COFECUB (\textit{Comit\'e
Fran\c{c}ais pour l'\'Evaluation de la Coop\'eration avec le Br\'esil}).

\end{document}